\documentclass[aps,prd,two column]{revtex4-2}

\usepackage{amsmath}
\usepackage{ulem}
\usepackage{color}
\usepackage{graphicx}
\usepackage{epsfig}
\usepackage{bm}
\usepackage{hyperref}
\usepackage{natbib}
\usepackage{caption}
\usepackage{subcaption}

\def\be {\begin{equation}}
\def\ee {\end{equation}}
\def\nn {\nonumber}
\def\bea {\begin{eqnarray}}
\def\eea {\end{eqnarray}}

\newcommand{\om}{\omega} 

\newcommand{\vk}{\vec k}

\begin{document}

\title{Thermoelectric figure of merit and the deconfinement phase transition}

\author{Kamaljeet Singh}
\author{Raghunath Sahoo}
\email{Corresponding Author: Raghunath.Sahoo@cern.ch}
\affiliation{Department of Physics, Indian Institute of Technology Indore, Simrol, Indore 453552, India}

\date{\today}
\begin{abstract}
Thermoelectric phenomena are traditionally associated with the interconversion of thermal and electrical energy in many-body systems. In the context of high-temperature quantum chromodynamics (QCD) matter produced in relativistic heavy-ion collisions, thermoelectric responses can provide insight into the evolving microscopic dynamics and the redistribution of effective degrees of freedom across the phase transition region. In this work, for the first time, we present a phenomenological study of the thermoelectric figure of merit (\( ZT \)) in hot QCD matter, with a particular focus on its behavior across the hadronic and quark-gluon plasma phases. Using model-based calculations for the electrical conductivity, Seebeck coefficient, and thermal conductivity, we analyze the temperature dependence of \( ZT \) and identify characteristic features near the QCD phase transition temperature. Our results indicate that \( ZT \) exhibits nontrivial behavior near the transition region, reflecting the changing transport properties and active degrees of freedom in the medium. This phenomenological study of the thermoelectric figure of merit provides a complementary perspective to traditional transport studies and may provide critical insights for advancing the understanding of QCD matter through the transition region.
\end{abstract}

\maketitle

\section{Introduction}
In the earliest moments of the universe after the Big Bang, approximately within the first few microseconds, the universe was supposed to exist in an extremely hot and dense state composed of deconfined quarks (antiquarks) and gluons. This primordial state of matter, known as the quark-gluon plasma (QGP)~\cite{Brewer:2019oha, Busza:2018rrf}, is remarkably a strongly interacting medium governed by the fundamental theory of strong interactions, Quantum Chromodynamics (QCD). With time, the universe underwent rapid expansion and cooling, leading to a phase transition where color charges became confined into hadrons~\cite{Lysenko:2024hqp}, a stage for the formation of matter as we observe it today. Direct access to the early universe is, of course, not possible. However, relativistic heavy-ion collisions at facilities such as the Large Hadron Collider (LHC) and the Relativistic Heavy Ion Collider (RHIC) provide a unique opportunity to recreate the extreme conditions of temperature and energy density that mimic the early universe scenario~\cite{Busza:2018rrf}. In these collisions, localized volumes of hot, deconfined state of quarks and gluons is formed, expanding and cooling rapidly, much like the cosmological QGP. This QGP medium exhibits strong collective flow~\cite{ Ambrus:2024eqa}, and nearly perfect fluidity~\cite{Bernhard:2019bmu}. These experiments provide deep insights into the thermodynamic~\cite{Pradhan:2023rvf,Khaidukov:2019icg, Koothottil:2018akg, Goswami:2023eol,Sahoo:2023vkw} and transport properties~\cite{Gavin:1985ph, K:2022pzc} of QCD matter. 
The transport coefficients associated with the QCD matter in both QGP and hadron resonance gas (HRG) phases, such as electrical conductivity~\cite{Singh:2023pwf,Dey:2020awu,Ghosh:2019ubc,Nam:2012sg,Hattori:2016cnt,Hattori:2016lqx,Harutyunyan:2016rxm,Kerbikov:2014ofa,Feng:2017tsh,Wang:2020qpx,Tuchin:2012mf}, thermal conductivity~\cite{Singh:2023pwf,Singh:2023ues}, charge diffusion coefficient~\cite{Goswami:2023hdl,Pradhan:2022gbm}, the thermoelectric Seebeck coefficient~\cite{Singh:2024emy,Li:2019bgc,PhysRevB.105.235116, Abhishek:2020wjm}, Thomson coefficient~\cite{Singh:2025rwc, Singh:2025geq}, and   shear and bulk viscosity~\cite{Huang:2011dc, Huang:2009ue,Agasian:2011st,Ghosh:2018cxb,Nam:2013fpa}, provide essential information about the microscopic off-equilibrium dynamics of QCD matter. 
 In conventional condensed matter systems, thermoelectric transport is primarily governed by a single dominant type of charge carrier, either electrons in n-type materials or holes in p-type materials. This carrier asymmetry allows a temperature gradient to induce a net electric current, giving rise to measurable thermoelectric effects such as the Seebeck and Thomson effects~\cite{book, PhysRev.98.940}. In contrast, an electron-ion plasma consists of two types of mobile charge carriers, electrons and comparatively heavier ions. While both species respond to temperature gradients, their opposite charges and differing mobilities can lead to competing contributions to the net electric current~\cite{ion}. In the ideal case of perfect symmetry or equal but opposite current responses, the thermoelectric contributions from electrons and ions may cancel each other. Therefore, in such plasmas, the generation of a net thermoelectric current typically requires additional asymmetries, such as mass or density gradients, or external fields beyond a mere temperature gradient. In the context of heavy-ion collisions, for the QGP phase, where quarks and gluons are deconfined, the transport of heat and charge is governed by partonic degrees of freedom, and thermoelectric effects, such as the Seebeck and Thomson coefficients, emerge due to the presence of temperature and chemical potential gradients generated in central to peripheral region of the created fireball~\cite{Singh:2025geq}. As the fireball cools and hadronizes, the system transitions into the HRG phase, where hadronic species dominate the dynamics. In this phase, thermoelectric transport is influenced by the spectrum of hadronic resonances and their scatterings, with contributions from charged mesons, baryons (antibaryons)~\cite{Singh:2025oja}.  

The thermoelectric efficiency of a medium is characterized by the dimensionless figure of merit, \( ZT = \frac{S^2 \sigma_{el} T}{\kappa_0} \), which combines the Seebeck coefficient (\( S \)), electrical conductivity (\( \sigma_{el} \)), and thermal conductivity (\( \kappa_0 \)) into a single quantity~\cite{Nemir2010, harpriya}. Here, \( \sigma_{el} \) determines how easily charge carriers can move through the material, contributing to electrical current. Whereas \( \kappa_0 \) quantifies the ability of a material to conduct heat, and \( S \) measures the magnitude of an induced thermoelectric voltage in response to a temperature gradient across the material.
Apart from estimating the thermoelectric efficiency of any material, $ZT$ also shows its sensitivity near the phase transition region. In many studies of condensed matter systems, the non-trivial behaviour of $ZT$ is observed near the phase transition region for different materials. In Ref.~\cite{10.1063/1.4870509}, a peak of $ZT$ is observed near the temperature of the phase transition from the semiconducting state to a superionic state of the material silver selenide ($Ag_2Se$). Similarly, in Ref.~\cite{10.1063/1.4827595}, a dramatic increase in $ZT$ near the phase transition in copper chalcogenides such as copper selenide ($Cu_2Se$).
Also, the peak of $ZT$ is observed near the transition temperature for various materials such as tin selenide ($SnSe$), zinc antimonide ($Zn_4Sb_3$), etc.~\cite{doi:10.1021/jacs.7b05881, LIN2021100787}. During the cubic-to-rhombohedral phase transition in germanium telluride ($GeTe$), the peak in $ZT$ value is observed~\cite{gete}. For polycrystalline high temperature superconductors such as mercury barium calcium copper oxide ($HgBaCaCuO$), bismuth strontium calcium copper oxide ($BiSrCaCuO$), and dysprosium barium copper oxide ($DyBaCuO$), the $ZT$ presents a remarkable peak for all at the critical temperature~\cite{bougrine1998figure}.
In the context of hot and dense QCD matter, \( ZT \) is not directly accessible in experiments; however, it serves as a valuable theoretical indicator of the interplay between charge and heat transport. Importantly, \( ZT \) is highly sensitive to changes in the microscopic structure of the medium, particularly near the QCD phase transition, where the underlying degrees of freedom and interaction dynamics evolve rapidly. In this study, we investigate the thermoelectric figure of merit of QCD matter across the hadronic and quark-gluon plasma phases. To study the QGP phase, we use a quasiparticle model (QPM)~\cite{Gorenstein:1995vm, Singh:2023ues, Singh:2025geq} where interactions are introduced through effective thermal masses of quarks and gluons. For the hadronic phase, we use the ideal Hadron Resonance Gas model~\cite{Das:2021qii, Singh:2025rwc, Singh:2025oja}, which successfully captures the thermodynamics of confined hadronic states. This two-phase framework allows us to analyze the temperature dependence of \( ZT \) and its sensitivity to the changing microscopic structure of the medium around the QCD phase transition region.


\section{Phenomenological models for the equation of state of QCD matter}\label{sec:formalism}
In this section, we discuss quasiparticle and hadron resonance gas models briefly. Further,
we present the calculations to evaluate the thermoelectric figure of merit for the QCD medium created in heavy-ion collisions. 

\subsection{Quasiparticle model}
\label{model1}
For a qualitative numerical description of QGP medium, we use a quasiparticle model proposed by Gorenstein and Yang~\cite{Gorenstein:1995vm} to study the QGP equation of state. 
In this phenomenological model, partons are considered with their thermal mass $m(T)$ arises from the interactions among the partons. The non-perturbative interactions of QGP medium are encoded into temperature-dependent effective masses and coupling constants, allowing a simplified description of the equation of state (EoS). Effective thermal masses are given to partons to reproduce the lattice QCD results, while a temperature-dependent bag constant is incorporated to ensure thermodynamic consistency by representing vacuum energy effects.
The energy \(\omega_i\) of a particle with momentum \(k_i\) satisfies the dispersion relation is $\om_i^{2}(k_i, T) = k_i^{2} + m_i^{2}(T)$. Where $m_i$ is the total effective mass of $i$th quark flavor and can be parameterized as 
\begin{align}
    m_i^2 = m_{i0}^2 + \sqrt{2}m_{i0}m_{iT} + m_{iT}^2~.
\end{align}
$m_{i0}$ and $m_{iT}$ represent the bare mass and thermal mass of the $i$th flavor, with
\begin{align}
    m_{iT}^2(T,\mu_B) = \frac{N_c^2 - 1}{8N_c}\Big( T^2 + \frac{\mu_B^2}{9 \pi^2}\Big)g^2(T, \mu_B).
\end{align}
In this model, the effective mass of the gluon $(m_g)$ can be represented as
\bea
m_{g}^2(T,\mu_B) = \frac{N_c}{6} g^2(T, \mu_B) T^2 \left(1+\frac{N_f + \frac{\mu_B^2}{\pi^2 T^2}}{6}\right).
\eea
$N_c$ represents the number of color degrees of freedom, and $g^2(T, \mu_B) = 4\pi \alpha_s(T, \mu_B)$, $\alpha_s(T, \mu_B)$ is running coupling constant. 
Now, for the relaxation time ($\tau_R$) of quarks, we use a momentum-independent expression obtained for QCD matter~\cite{Hosoya:1983xm}
\begin{align}
    \tau_R = \frac{1}{5.1T\alpha_s^{2}(T, \mu_B)\log\{1/\alpha_s(T, \mu_B)\}[1+0.12(2N_f+1)]}.
\end{align}
Here, in above calculations, we have considered the two-loop QCD coupling constant as~\cite{Srivastava:2010xa}
\begin{align}
&\alpha_s(T, \mu_B)
= \frac{6\pi}{
(33 - 2N_f)\,\ln\!\left(\frac{T}{\Lambda_T}
\sqrt{1 + \,\mu_B^2/9\pi^2T^2}\right)
} \times\nn \\
&\left[
1 - 
\frac{3(153 - 19 N_f)}{(33 - 2N_f)^2}
\frac{\ln\!\left(2\,\ln\!\left(\frac{T}{\Lambda_T}
\sqrt{1 + \,\mu_B^2/9\pi^2T^2}\right)\right)}{
\ln\!\left(\frac{T}{\Lambda_T}\sqrt{1 + \,\mu_B^2/9\pi^2T^2}\right)}
\right],\nonumber
\end{align}
where \(\Lambda_T\) is the QCD scale parameter. We take \(\Lambda_T =0.115\ \mathrm{GeV}\).
\subsection{Ideal hadron resonance model}
\label{model2}
The Ideal Hadron Resonance Gas model is a statistical framework used to describe strongly interacting matter in the hadronic phase, mainly at temperatures below the QCD crossover transition~\cite{Singh:2025oja}. In this model, the thermodynamic properties of the medium are calculated under the assumption that it consists of a non-interacting gas of all known hadrons and resonances listed in the Particle Data Group (PDG)~\cite{ParticleDataGroup:2008zun}.
Each hadronic species contributes independently to the total thermodynamic quantities, using the Fermi-Dirac or Bose-Einstein statistics, depending on whether the particle is a baryon or meson. The inclusion of resonances effectively mimics the strong interactions among hadrons based on the idea that interactions in the medium can be approximated by resonance formation. For a system with volume $V$ having non-interacting pointlike hadrons and resonances, for which the grand canonical partition function can be written as~\cite{Pradhan:2022gbm},
\begin{equation}
\label{eq1}
ln Z^{id}_i = \pm \frac{Vg_i}{2\pi^2} \int_{0}^{\infty} k_i^2 dk_i\ ln\{1\pm \exp[-(\omega_i-\mu_i)/T]\}.
\end{equation}
Here, $g_i$ is the degeneracy factor, and the quantities $k_i$, $m_i$, and $\omega_i = \sqrt{k_i^2 + m_i^2}$ represents the momentum, mass, and energy of the $i{\rm th}$ hadron species, respectively. The $\pm$ sign corresponds to baryons (upper) and mesons (lower). Considering a simplistic case of vanishing charge and strangeness chemical potential, the total chemical potential of the $i$th hadronic species, $\mu_{i}$, is equal to the baryon chemical potential ($\mu_B$) and is given by 
\begin{equation}
\label{eq2}
\mu_i = b_i\mu_B ,
\end{equation}
where $b_i$ denotes the baryon number of the $i{\rm th}$ hadron. The pressure $P_i$, energy density $\varepsilon_i$, and number density $n_i$ can now be obtained from the partition function, given as,
\begin{align}
\label{eq3}
P^{id}_i(T,\mu_i) &= \pm \frac{Tg_i}{2\pi^2} \int_{0}^{\infty} k_i^2 dk_i\ ln\{1\pm \exp[-(\omega_i-\mu_i)/T]\}, \\
\varepsilon^{id}_i(T,\mu_i) &= \frac{g_i}{2\pi^2} \int_{0}^{\infty} \frac{\omega_i\  k_i^2 dk_i}{\exp[(\omega_i-\mu_i)/T]\pm1},\\
n^{id}_i(T,\mu_i) &= \frac{g_i}{2\pi^2} \int_{0}^{\infty} \frac{k_i^2 dk_i}{\exp[(\omega_i-\mu_i)/T]\pm1}.
\end{align}
For the case of a hadronic medium, we take the thermal-averaged relaxation time after integrating the energy-dependent relaxation time over the equilibrium distribution function. The thermal averaged relaxation time ($\tau_{R}^i$) for the $i$th hadron species can be expressed in terms of scattering cross-section as~\cite{Das:2020beh},
\begin{align}
    {\tau_{R}^i}^{-1} = \sum_j n_j \langle\sigma_{ij}v_{ij}\rangle
\end{align}
where,
\begin{align}
 \langle\sigma_{ij}v_{ij}\rangle &= \frac{\sigma}{8Tm_i^2m_j^2\mathcal{K}_2(m_i/T)\mathcal{K}_2(m_j/T)}~ \times\nn\\
 &\int_{(m_i+m_j)^2}^{\infty}ds~\times \nn\\
 &\frac{[s-(m_i-m_j)^2]}{\sqrt{s}}
 \times [s-(m_i+m_j)^2]\mathcal{K}_1(\sqrt{s}/T),
 \label{equnew70}
\end{align}
 where the total scattering cross-section for the hard spheres is $\sigma = 4\pi r_h^2$, which is independent of both temperature and baryon chemical potential. $\mathcal{K}_1, \mathcal{K}_2$ are modified Bessel functions of the first and second order.

\section{Thermoelectric figure of merit of QCD matter}
\label{formalism2}
 To calculate the thermoelectric figure of merit of QCD matter, we follow a kinetic theory approach. 
 We first consider the total single-particle distribution function $f_i=f^0_i+\delta f_i$. Here, $f_i^{0}$ is the single-particle equilibrium distribution function and $\delta f_i$ represents the deviation from the equilibrium state.  The total single-particle distribution function for $i{\rm th}$ species at equilibrium is given by
\begin{align}\label{Dis-f}
f^0_i= \frac{1}{e^{\frac{\omega_i-{\rm b}_i\mu_B}{T}}\pm 1}~,
\end{align}
where $\omega_i$ is the single particle energy,  $\mu_B$ is the baryon chemical potential, $b_i$ denotes the baryon number of $i{\rm th}$ species, e.g. for baryons $b_i=1$, for anti-baryons $b_i=-1$ and for mesons $b_i=0$. The $\pm$ sign stands for fermions and bosons, respectively. Now, we can write the linearized BTE under RTA in the local rest frame (LRF), for particle species $i$ as \cite{Das:2020beh, Singh:2023pwf},

\begin{equation}
\frac{\partial f_i}{\partial t} + \vec{v}_i.\vec{\nabla}f_i+q_i\vec{E}.\frac{\partial f_i}{\partial\vec{k_i}} = 
  -\frac{\delta f_i(\vec{x_i},\vec{k_i})}{\tau^{i}_R} ,
 \label{equnew1}
\end{equation}
where $\vec{E}$ is the non-zero electric field that drives the system out of thermal equilibrium and  $\tau^{i}_R$ denotes the relaxation time of the particle species $i$. The equilibrium distribution function satisfies, 
\begin{align}
 \frac{\partial f^{0}_{i}}{\partial \vec{k_i}}=\vec{v}_i\frac{\partial f^{0}_{i}}{\partial \omega_i},
 ~~ \frac{\partial f^{0}_{i}}{\partial \omega_i}=-\frac{f^{0}_{i}(1\mp f^{0}_{i})}{T}, 
\label{equnew2}
 \end{align}
  $\vec{v}_i=\vec{k_i}/\omega_i$ is the velocity of the $i$th particle. The gradient of the equilibrium distribution function $\vec{\nabla}f^{0}_{i}$ can be expressed as,
\begin{align}
 \vec{\nabla}f^{0}_{i} = T \bigg[\omega_i\vec{\nabla}\left(\frac{1}{T}\right)-b_i\vec{\nabla}\left(\frac{\mu_B}{T}\right)\bigg]\frac{\partial f^{0}_{i}}{\partial\omega_i}.
 \label{equnew3}
\end{align}

Using the Gibbs-Duhem relation, we then have,
\begin{align}
 \vec{\nabla}f^{0}_{i} = - \frac{\partial f^{0}_{i}}{\partial\omega_i}\bigg(\omega_i-b_i h\bigg)\frac{\vec{\nabla}T}{T}.
\label{equnew4}
 \end{align}
 where $h=\frac{\varepsilon + P}{n}$ is the enthalpy per particle, $\varepsilon$, $P$, and $n$ are total energy density, total pressure, and net baryon density of the system, respectively.
 With leading order contributions, we can write an ansatz of $\delta f_i$ as~\cite{Gavin:1985ph, Singh:2023pwf, Das:2019pqd}
\begin{align}\label{delta-f0}
	\delta f_i = (\vec{k_i} \cdot \vec{\Omega}) \frac{\partial f^0_i}{\partial \omega_i}~.
\end{align}  
In general, the preferred form of unknown vector ${\vec \Omega}$ can be assumed as a linear combination of all existing perturbing forces leading the system out of thermal equilibrium as \begin{align}\label{Omega00}
\vec{\Omega} = &~\alpha_1 \vec{E} +\alpha_2 \vec{\nabla}T ~.
\end{align}  
The unknown coefficients $\alpha_j$ ($j=1,2$) determine the strength of the respective gradient force fields driving the system away from equilibrium.
Using Eq.\eqref{equnew4} and Eq.\eqref{equnew2} in Eq.\eqref{equnew1}, we can write the deviation of the equilibrium distribution function as,
\begin{align}
 \delta f_i = -\tau^{i}_R\frac{\partial f_i^{0}}{\partial \omega_i}\bigg[q_i(\vec{E}.\vec{v}_i)-\bigg(\frac{\omega_i-b_i h}{T}\bigg)\vec{v}_i.\vec{\nabla} T\bigg].
 \label{equnew5}
\end{align}
Following the kinetic theory, the electric current $(\vec{j})$ of the system can be written in terms of the deviation from the equilibrium distribution function $\delta f_i$ as,
\begin{align}
 \vec{j}& = \sum_i g_i\int \frac{d^3|\vk_i|}{(2\pi)^3}q_i\vec{v}_i\delta f_i\nonumber\\
 & = \sum_i \frac{g_i}{3} \int \frac{d^3|\vk_i|}{(2\pi)^3}\tau^{i}_R q_i^2 v^2_i\bigg(-\frac{\partial f_i^{0}}{\partial \omega_i}\bigg)\vec{E}\nonumber\\
 & -\sum_i \frac{g_i}{3} \int \frac{d^3|\vk_i|}{(2\pi)^3}\tau^{i}_Rq_iv^2_i
 \bigg(\frac{\omega_i-b_i h}{T}\bigg)\bigg(-\frac{\partial f_i^{0}}{\partial \omega_i}\bigg)\vec{\nabla} T.
 \label{equnew6}
\end{align}
In the above equation, we have used $\langle v^l_iv^j_i\rangle=\frac{1}{3}v_i^2\delta^{lj}$. Here, the sum is over all the quarks and anti-quarks. 
For a relativistic system, one can also define the thermal current with reference to the conserved baryon current. The thermal current arises when energy flows relative to the baryonic enthalpy. 
Hence, the heat current of the QGP medium can be defined as \cite{Gavin:1985ph},

\begin{align}
\label{heat_current}
 \vec{I} & = \sum_i g_i\int \frac{d^3|\vk_i|}{(2\pi)^3}\frac{\vec{k}_i}{\omega_i}\left(\omega_i-b_i h\right)\delta f_i.\nonumber\\
& = \sum_i \frac{g_i}{3} \int \frac{d^3|\vk_i|}{(2\pi)^3} \tau^{i}_R q_i v_i^2\left(\omega_i-b_i h\right)\left(-\frac{\partial f_i^{0}}{\partial\omega_i}\right)\vec{E}\nonumber\\
 & -\sum_i \frac{g_i}{3T} \int \frac{d^3|\vk_i|}{(2\pi)^3} \tau^{i}_R v_i^2\left(\omega_i-b_i h\right)^2\left(-\frac{\partial f_i^{0}}{\partial\omega_i}\right)\vec{\nabla}T.
\end{align}
One can define the Seebeck coefficient $S$ using Eq.\eqref{equnew6} by setting $\vec{j}=0$ such that the electric field and temperature gradient become proportional to each other. Here, the proportionality factor is known as the Seebeck coefficient \cite{Singh:2024emy}. Hence from  Eq.\eqref{equnew6} we get, 
\begin{align}
 \vec{E}=S\vec{\nabla}T,
\end{align}
hence, 
\begin{align}
 S &= \frac{\sum_i \frac{g_i}{3}\int \frac{d^3|\vk_i|}{(2\pi)^3}\tau^{i}_R q_i v_i^2\left(\omega_i-b_ih\right)\left(-\frac{\partial f_i^{0}}{\partial\omega_i}\right)}{T\sum_i \frac{g_i}{3}\int \frac{d^3|\vk_i|}{(2\pi)^3}\tau^{i}_R q^2_i v_i^2\left(-\frac{\partial f_i^{0}}{\partial\omega_i}\right)}\nonumber\\
 &=\frac{\sum_i \frac{g_i}{3T}\int \frac{d^3|\vk_i|}{(2\pi)^3}\tau^{i}_R q_i \left(\frac{\vec{k_i}}{\omega_i}\right)^2\left(\omega_i-b_ih\right)f^{0}_{i}(1\mp f^{0}_{i})}{T\sum_i \frac{g_i}{3T}\int \frac{d^3|\vk_i|}{(2\pi)^3}\tau^{i}_R q^2_i \left(\frac{\vec{k_i}}{\omega_i}\right)^2f^{0}_{i}(1\mp f^{0}_{i})}\nonumber\\
 &= \frac{\mathcal{I}_{1}/T^2}{\sigma_{el}/T}.
 \label{equnew11}
\end{align}
Where, the electrical conductivity ($\sigma_{el}$) can be identified from Eq.\eqref{equnew6} as,
\begin{align}
 \sigma_{el}  & = \sum_i \frac{g_i}{3T}\int \frac{d^3|\vk_i|}{(2\pi)^3}\tau^{i}_R q^2_i \left(\frac{\vec{k_i}}{\omega_i}\right)^2f^{0}_{i}(1\mp f^{0}_{i}),
\end{align}
and the integral $\mathcal{I}_{1}$ in Eq.\eqref{equnew11} is, 
\begin{align}
 \mathcal{I}_{1} = \sum_i \frac{g_i}{3T}\int \frac{d^3|\vk_i|}{(2\pi)^3}\tau^{i}_R q_i \left(\frac{\vec{k_i}}{\omega_i}\right)^2\left(\omega_i-b_ih\right)f^{0}_{i}(1\mp f^{0}_{i}). 
 \label{I31equ}
\end{align}
The mesons contribute through the total enthalpy of the system as well as in the total electrical conductivity of the system, which enters the denominator of Eq. \eqref {equnew11}. 
It is to be noted that the Seebeck coefficient can be both positive and negative because the numerator depends linearly on an electric charge while the integrand itself is not positive definite. The electric current and heat current can be modified due to these thermoelectric coefficients as,
\begin{align}
 \vec{j}&=\sigma_{el}\vec{E}-\sigma_{el}S \vec{\nabla}T. \label{equnew14a}\\
 \vec{{I}}&=T\sigma_{el}S\vec{E}-\kappa_0\vec{\nabla}T,
 \label{equnew14}
\end{align}
where $\kappa_0$ is the coefficient of the thermal conductivity and is expressed as \cite{Singh:2023pwf},
\begin{align}
 \kappa_0=\sum_i\frac{g_i}{3T^2}\int\frac{d^3|\vk_i|}{(2\pi)^3}\tau^{i}_R\left(\frac{\vec{k_i}}{\omega_i}\right)^2\left(\omega_i-b_ih\right)^2f^{0}_{i}(1\mp f^{0}_{i}).
 \label{equnew15}
\end{align}
Using Eq.\eqref{equnew14a} and Eq.\eqref{equnew14}, we can express the heat current $\vec{{I}}$ in terms  of electric current $\vec{j}$ in the following way,
\begin{align}
 \vec{{I}}&=TS\vec{j}-\left(\kappa_0-S^{2}\sigma_{el}T\right)\vec{\nabla}T, \nonumber\\
 &=TS\vec{j}-\tilde \kappa_0\vec{\nabla}T, \nonumber\\
          &=TS\vec{j}-\left(1-\frac{S^{2}\sigma_{el}T}{\kappa_0}\right)\kappa_0\vec{\nabla}T.
 \label{equnew16}
\end{align}
 Here, $\tilde \kappa_0 = \kappa_0-S^{2}\sigma_{el}T$ represents the coefficient of effective thermal conductivity in presence of the electric field. The three transport coefficients, namely $S$, $\sigma_{el}$, and $\kappa_0$, are closely related to each other because of the common factors such as mobility and concentration of medium constituents. The thermoelectric performance of any thermoelectric material can be measured by using the dimensionless quantity named as the figure of merit ($ZT$), given as~\cite{Nemir2010}
\begin{align}
\label{eqnZT}
    ZT = \frac{S^{2}\sigma_{el}T}{\kappa_{0}}.
\end{align}
 Here, the term $S^{2}\sigma_{el}$ in the numerator of the above equation is called the power factor. It reflects the ability of the thermoelectric medium to convert heat to electrical energy, excluding the effects of heat conduction. The first term in Eq.~\eqref{equnew16} represents the heat carried by charge carriers due to the Seebeck effect. When an electric current $\vec{j}$ flows in a material with a non-zero Seebeck coefficient, the carriers transport energy, thus producing a heat current proportional to $TS\vec{j}$. Whereas the second term describes heat flow due to the temperature gradient, similar to Fourier's law, but corrected by the factor $(1-ZT)$. This correction shows that the presence of charge carriers affects the net heat flow.
Hence, the systems having $ZT$ = 0 behave like a regular thermal conductor, and heat flows in response to the temperature gradient with no thermoelectric effect. Whereas, for the systems having $ZT \neq$ 0, a portion of the temperature-driven heat flow is converted into electric energy or compensated by the thermoelectric current. 
\begin{figure*}
	\centering
    \raisebox{10mm}{\includegraphics[scale=0.40]{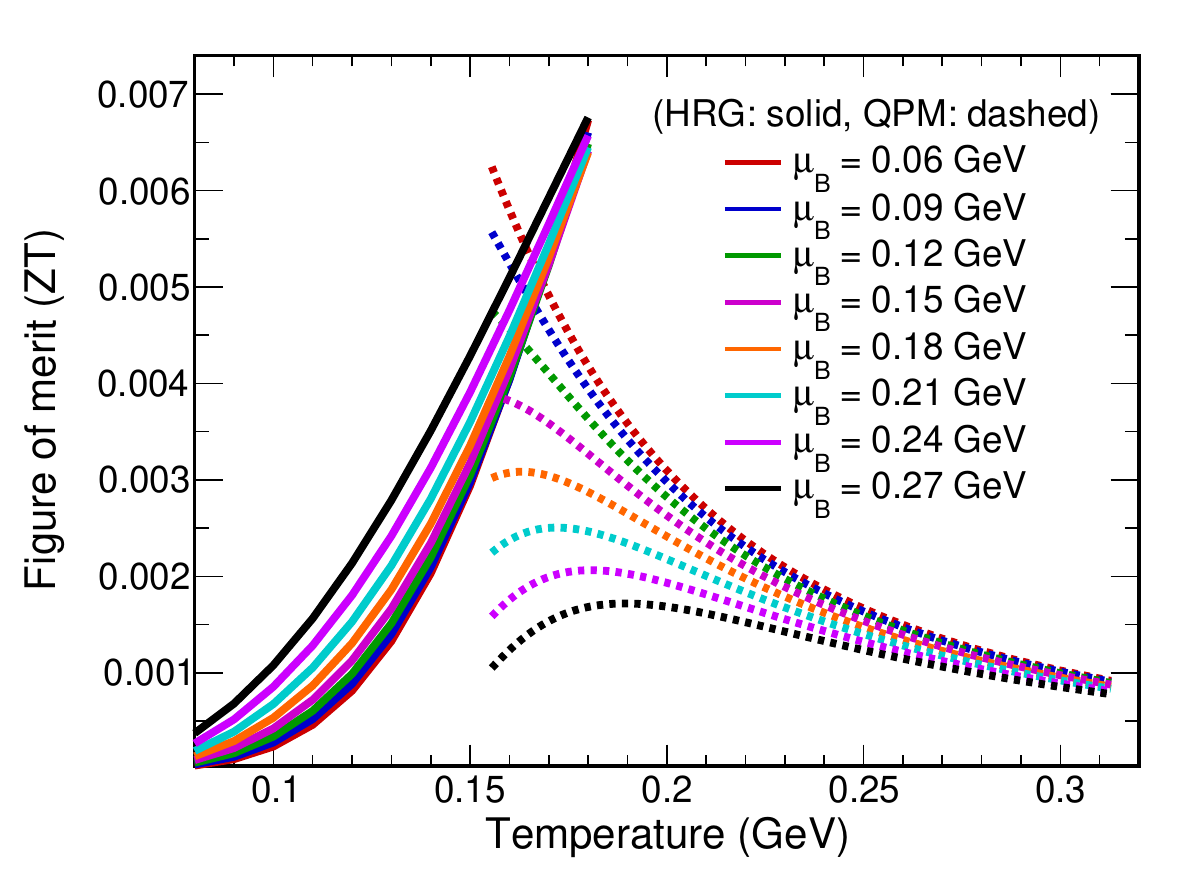}}
    \includegraphics[scale=0.40]{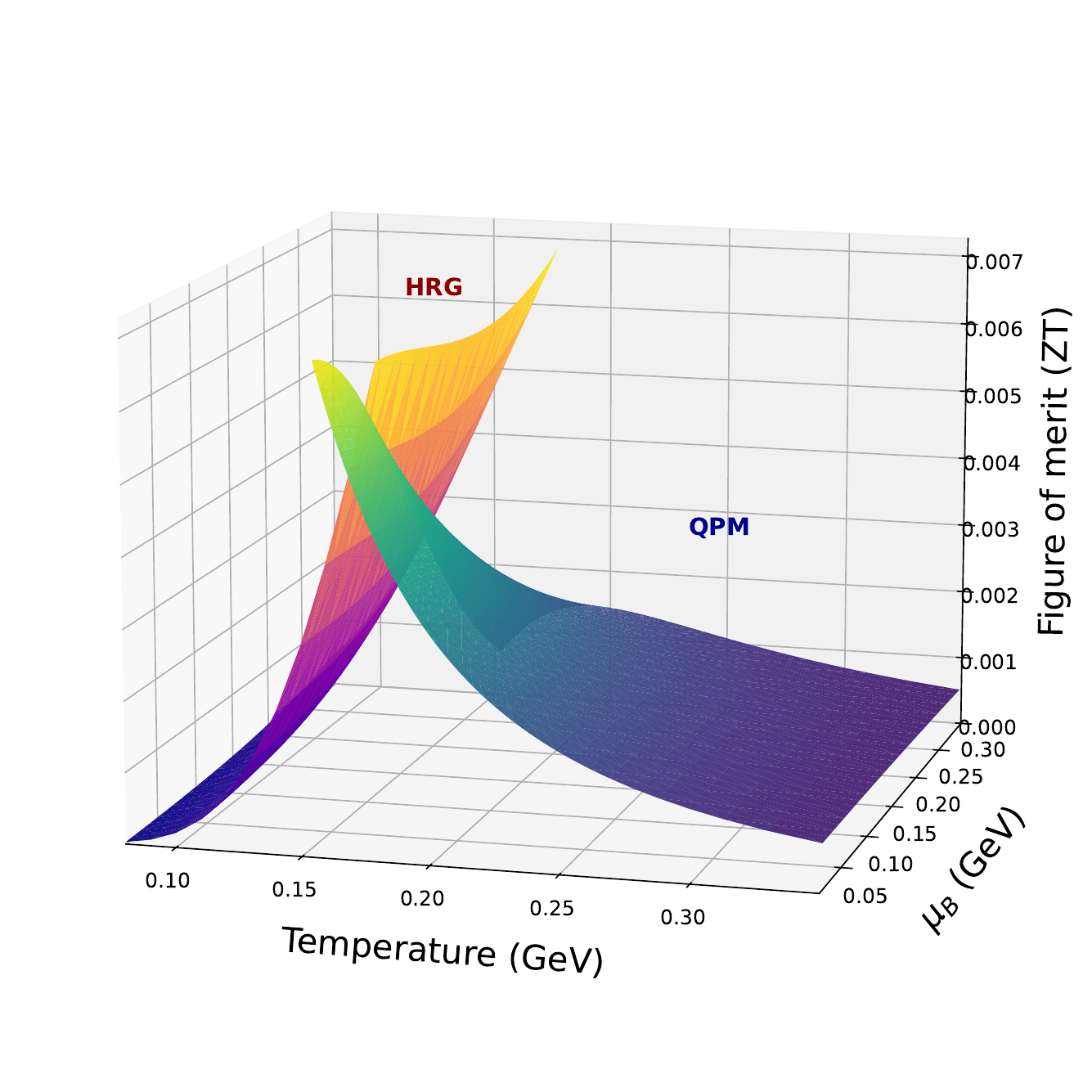}
	 \caption{Left: $ZT$ as a function of temperature for various values of $\mu_B$ in the HRG (solid lines) and QPM (dashed lines) phases. Right: 3D surface plot showing $ZT$ across a thermodynamic phase space spanned by $T$ and $\mu_B$ in both HRG and QPM phases.}
    \label{fig:ZT_plots}
\end{figure*}

\section{Results and Discussion}\label{Sec-results}
For the very first time, we have calculated the thermoelectric figure of merit \( ZT \) for QCD matter created in heavy-ion collisions. As mentioned in Eq.\eqref{eqnZT}, \( ZT \) serves as a key dimensionless parameter to quantify the efficiency of a medium in converting a temperature gradient into electrical energy.  The Seebeck coefficient \( S \) characterizes the voltage induced per unit temperature gradient across the medium and reflects the strength of the thermoelectric response. The electrical conductivity \( \sigma_{el} \) measures the ability of the medium to conduct electric current, while the thermal conductivity \( \kappa_0 \), comprising both baryonic and mesonic contributions, governs the transport of heat. A high value of \( ZT \) indicates an efficient thermoelectric material, which ideally requires a large Seebeck coefficient, high electrical conductivity, and low thermal conductivity. In this study, we explore the temperature dependence of $ZT$ for various values of baryon chemical potential in both the HRG medium and QGP medium. Below, without repeating all the details, we discuss the behavior of each transport coefficient first and then discuss their combined impact on $ZT$ in detail. 

The Seebeck coefficient $S$ shows a strong temperature and $\mu_B$ dependence. In the HRG phase, $S$ becomes increasingly negative with temperature as mentioned in our previous study~\cite{Singh:2025rwc}. This reflects enhanced baryonic transport and increasing charge asymmetry. In relativistic systems, defining heat flow requires a conserved charge. For the HRG medium, this role is played by the net baryon number, making the net heat current directly related to the baryon current. Here, baryons are found to contribute more significantly to the Seebeck coefficient than mesons. The contribution of mesons to thermoelectric properties arises solely through the enthalpy of the medium. Notably, light mesons such as pions and kaons contribute largely to entropy production, which in turn elevates the enthalpy per baryon ($h$) over the single-particle energy ($\omega_i$) in Eq.~\ref{equnew11}. This imbalance is responsible for the emergence of negative Seebeck coefficients in the HRG medium. In contrast, the QGP phase exhibits charged quarks and neutral gluons. The quarks have the leading contribution to thermoelectric transport of the medium, whereas gluons contribute through the enthalpy. For this phase, the $S$ is observed to be almost saturated in the higher temperature region~\cite{Singh:2025geq}. This distinction in $S$ behavior between hadronic and deconfined phases critically influences the temperature profile of $ZT$ through the $S^2$ term. While the sign of the Seebeck coefficient indicates the relative alignment between the electric field and the temperature gradient, this directional information is lost in the $ZT$, where $S$ appears squared and thus contributes only through its magnitude. The numerator of Eq.~\ref{equnew11} also shows the dependence of an electrical conductivity $\sigma_{el}$ of the medium. As it is proportional to electric charge squared, it is positive for both phases. In the hadronic phase, $\sigma_{el}$ decreases with temperature and increasing values of $\mu_B$ mainly due to the decrease of the relaxation time of the mesons. However, in the QGP phase, the $\sigma_{el}$ increases with both $T$ and $\mu_B$ because of an increase in the relaxation time of quarks. This turnaround in both phases, HRG and QPM, contributes to shaping the peak structure in $ZT$. In the denominator of Eq.~\ref{equnew11}, the thermal conductivity $\kappa_0$ decreases with $T$ and $\mu_B$ in the HRG phase~\cite{Singh:2023pwf}, but on the contrary, it increases in the QGP phase as $T$ increases but decreases with $\mu_B$~\cite{Singh:2023ues}. As mentioned in Eq.~\ref{equnew15}, the $\kappa_0$ has a contribution of the square of $(\omega_i-b_ih)$, hence it is positively contributing to $ZT$. Therefore, a dip near the transition region contributes to $ZT$ inversely. The combined effect of $S$, $\sigma_{el}$, and $\kappa_0$ results in a clear peak in $ZT$ near the QCD crossover temperature region. This peak signifies the region where thermoelectric conversion is most efficient. In a nutshell, the thermoelectric figure of merit is a powerful combined measure of transport efficiency, capturing the interplay between electric and thermal responses in a single, dimensionless quantity. Unlike the individual transport coefficients, $ZT$ integrates their effects to reflect the overall ability of a system to convert thermal gradients into electrical energy. Its strength lies in its sensitivity to collective medium properties, making it especially valuable near the phase transition, where rapid changes in transport dynamics occur. A peak in this combined quantity can signal critical behavior more effectively than any single transport coefficient on its own.

Figure~\ref{fig:ZT_plots} displays the behavior of the thermoelectric figure of merit $ZT$ as a function of temperature $T$ and baryon chemical potential $\mu_B$, for the hadronic phase using the HRG model, and the QGP phase using the quasiparticle model. The left panel shows 2D profiles of $ZT$ versus $T$ for fixed values of $\mu_B$, while the right panel presents the full 3D surface of $ZT$ across a thermodynamic phase space spanned by both $T$ and $\mu_B$.
In the left panel, solid lines correspond to the HRG model, and dashed lines represent the QPM. The $ZT$ increases monotonically with temperature within the HRG model up to the transition region, after which it sharply drops for the QGP medium described by the quasi-particle model. For low $\mu_B$ values (e.g., $\mu_B = 0.06$ GeV), the peak in $ZT$ is smooth around the transition temperature region. As $\mu_B$ increases, the discontinuity starts appearing in the peak of both phases. Notably, at high temperatures, $ZT$ is nearly independent of $\mu_B$ in the QPM, which may be due to the existence of near-symmetry of quark and antiquark contributions and the negligible role of net baryon number in a weakly interacting quark-gluon plasma.
The right panel of Fig.~\ref{fig:ZT_plots} further highlights the 3D surface plot of $ZT$ with respect to $T$ and $\mu_B$. In the HRG model, $ZT$ grows significantly with both $T$ and $\mu_B$ up to the transition region, peaking sharply before dropping in the QGP region. These results collectively suggest that the thermoelectric figure of merit could serve as a sensitive probe of the QCD transition. The presence of a peak in $ZT$ near the transition may provide a novel, transport-based phenomenological signature of the deconfinement transition and the evolution of charge-carrying degrees of freedom in hot and dense QCD matter. 
It is important to note here that the point of discontinuity in the temperature dependence of $ZT$ occurs around $T \simeq$ 160 MeV for lower values of $\mu_B$. This corresponds to the lattice QCD predicted value of a deconfinement phase transition \cite{HotQCD:2018pds}.

\section{Summary}\label{sec-summary}
In summary, for the first time, we have calculated the thermoelectric figure of merit \( ZT \) of the QCD medium. Thermoelectric transport represents a fundamental and elegant interplay between heat and charge flow in a medium. In condensed matter systems, it forms the foundation of technologies that enable direct conversion between thermal and electrical energy, presenting promising opportunities for energy harvesting and thermal management. The study of \( ZT \) plays a crucial role in designing efficient materials for these technologies. Beyond its practical relevance, \( ZT \) serves as a sensitive probe of the microscopic structure and carrier dynamics of a material. It shows how charge carriers respond to temperature gradients and how thermal excitations influence electrical conduction. This dual sensitivity provides deeper insight into the intrinsic properties of diverse systems, ranging from conventional solids to strongly interacting quantum matter, making thermoelectric phenomena not only technologically valuable but also fundamentally rich in exploring the transport behavior of complex media. In our current study, we observe that around the QCD phase transition region, \( ZT \) exhibits nontrivial behavior due to the rapid changes in the transport coefficients. Near the transition temperature region, the Seebeck coefficient \( S \), \( \sigma_{el} \), and  \( \kappa_0 \) undergo significant modifications due to the change in degrees of freedom from hadronic to partonic medium. These combined effects can lead to a pronounced peak in \( ZT \), suggesting that the medium is most thermoelectrically responsive near the phase transition region. We have also observed that for the lower values of $\mu_B$, the peaks of $ZT$ for both regions overlap around the transition region. However, with increasing baryon asymmetry, the discontinuity between peaks of $ZT$ corresponding to both phases also increases. Hence, the sensitivity of $ZT$ reflects its valuable tendency to understand the complex dynamics around the phase transition region of baryon-free to baryon-rich QCD matter in the context of heavy-ion collisions.  
 
\acknowledgments
K.S. acknowledges the doctoral fellowship from the UGC, Government of India. R.S. gratefully acknowledges the DAE-DST, Govt. of India funding under the mega-science project – “Indian participation in the ALICE experiment at CERN” bearing Project No. SR/MF/PS-02/2021-IITI (E-37123). The authors would like to thank Suraj Prasad for careful reading of the paper and providing valuable feedback.

\end{document}